\def\year{2017}\relax
\documentclass[letterpaper]{article}
\usepackage{aaai17}
\usepackage{times}
\usepackage{helvet}
\usepackage{courier}
\usepackage[utf8]{inputenc}
\usepackage{subfig}
\usepackage{url} 
\usepackage{booktabs}
\usepackage{CJKutf8}
\usepackage{graphicx}
\usepackage{amsmath}
\usepackage{multirow}
\usepackage{nameref}

\newcommand{\dress}[1]{{``The Dress''}{}}
\newcommand{\demo}[1]{{\url{http://ciir.cs.umass.edu/irdemo/contention/}}{}}

\title{Is Climate Change Controversial? Modeling Controversy as Contention Within Populations}

 \author{
 Shiri Dori-Hacohen, Myungha Jang and James Allan \\ Center for Intelligent Information Retrieval \\ College of Information and Computer Sciences \\ University of Massachusetts}

\begin{document}


\maketitle

\begin{abstract}
A growing body of research focuses on computationally detecting controversial topics and understanding the stances people hold on them. Yet gaps remain in our theoretical and practical understanding of how to define controversy, how it manifests, and how to measure it. In this paper, we introduce a novel measure we call ``contention'', defined with respect to a topic and a population. We model contention from a mathematical standpoint. We validate our model by examining a diverse set of sources: real-world polling data sets, actual voter data, and Twitter coverage on several topics. In our publicly-released Twitter data set of nearly 100M tweets, we examine several topics such as Brexit, the 2016 U.S. Elections, and ``The Dress'', and cross-reference them with other sources. We demonstrate that the contention measure holds explanatory power for a wide variety of observed phenomena, such as controversies over climate change and other topics that are well within scientific consensus. Finally, we re-examine the notion of controversy, and present a theoretical framework that defines it in terms of population. We present preliminary evidence suggesting that contention is one dimension of controversy, along with others, such as ``importance''. Our new contention measure, along with the hypothesized model of controversy, suggest several avenues for future work in this emerging interdisciplinary research area.
\end{abstract}








\vspace*{-1mm}
\section{Introduction}

Social network tools such as Twitter, Facebook, discussion forums, and comments on news articles 
are increasingly the place where democratic arguments are being held. Technological tools hold an increasingly crucial role in shaping these discussions by influencing which users see which data, through algorithmic curation and filtering. 
The current state of affairs is that we simply do not understand controversy well enough from a computational perspective. Algorithms based on incomplete understanding are bound to fail in a variety of unexpected ways, replicating or even exacerbating the sources of human bias in the data. 

Recent work on controversy cuts across traditional disciplinary lines to include a wide variety of computational tasks along with social science and humanities \cite{Dori-HacohenPosition}, and has made significant strides in analyzing and detecting controversy (cf. \cite{Garimella2016,Borra2015}). Nonetheless, serious gaps remain in our theoretical and practical understanding of how to define controversy, and how it manifests and evolves. 
For example, polling organizations naturally segment their results based on population groups such as race and gender, but these notions are surprisingly absent from algorithmic analyses of online data. Instead, controversy is assumed to be an absolute, single value for an amorphous global population.

Meanwhile, a disparity is growing between scientific understanding and public opinion on certain controversial topics, such as climate change, evolution, or vaccines \cite{Leshner2015}, with many scientists explicitly fighting these trends by insisting ``there is no controversy'' \cite{helfand2016survival} (referring to \textit{scientific} controversy). Still, non-scientific claims and arguments continue to proliferate, raising exposure to the (supposedly non-existent) controversies. As researchers studying controversies online, how are we to reconcile the oft-repeated argument from the scientific community that ``there is no controversy'' with the practical appearance of wildly diverse opinions on said topics? In other words, is climate change controversial\footnote{This differs from a value judgment, such as ``Should climate change be controversial?''.}?


This paper addresses these issues by proposing a theoretical framework for a new measure we call ``contention'', drawing on insights from social science and humanities and marrying them with the mathematical rigor of a computational approach. Our framework departs from most existing work about controversy in a two major ways. First, we define contention not only in terms of its topic, but also in terms of the population being observed. Second, our model accounts for participants in the population who hold no stance with regards to a specific topic, and also allows for any number of stances rather than just two opinions. These elements give our model explanatory power that can be used to understand a large variety of observed phenomena, ranging from international conflict, through community-specific controversies, as well as the aforementioned high-stakes public opinion controversies over scientifically well-understood phenomena such as climate change, evolution, and vaccines. 

In order to ground our theoretical model, we examine a diverse collection of data sets from both online and offline sources. First, we examine several real-world polling data sets, among them a poll that focuses on opinions about scientific topics, such as climate change and evolution, measured among the general U.S. population as well as the scientific community \cite{PewResearchCenter2015,PewResearchCenter2015a}. Additionally, we look at Twitter coverage for several popular topics in the last eighteen months, including three prominent controversies (the 2016 U.S. Elections, the UK referendum on leaving the EU, commonly known as Brexit, and \dress{}, a photo that went viral when people disagreed on its colors). We cross-reference contention from Twitter with other data sources: a popular online poll for \dress{}, and actual voter data for Brexit and the U.S. Elections.

Finally, we reexamine the concept of controversy in light of the new population-based contention measure as well as the case studies chosen. We hypothesize a general model of controversy as composed of at least two dimensions, rather than being a one-dimensional quantity, and propose that like our contention measure, it is population-dependent. We present preliminary evidence to support this model, which contains contention as one of its dimensions, and hypothesize ``importance'' as a minimal additional dimension.

Our new contention measure, along with the hypothesized model of controversy, afford new directions of understanding of controversy, such as the growth of contention or controversy over time among different populations, and points to open questions for future research.

\section{Prior Work}
\label{sec:priorwork}

Research on controversies in computer science has nearly universally considered controversy as either a binary state or a single quantity, both of which are to be measured or estimated directly \cite{Awadallah2012,RadComparative,Borra2015}. With few exceptions \cite{Amendola2015,Jang2016a}, prior work did not model controversy formally. Even when it did, the meaning of controversy was not modeled, but assumed to be a known quantity in the world. Most prior work in computer science does not define controversy at all, and treats it as a global quantity (cf. \cite{Kittur,Yasseri2013}). We refer the interested reader to a recent survey of the field which discusses the problem of defining controversy, which is a complex problem \cite{Dori-HacohenPosition}. Past research shows that achieving inter-annotator agreement on the ``controversy'' label is challenging \cite{Dori-Hacohen,Klenner14}. We depart from prior work in computer science by focusing on a more achievable goal of measuring what we call ``contention'', a population-dependent measure, and offering a mathematical framework to define it while grounding it in empirical data.


Meanwhile, most of the work on controversy in social studies and humanities is qualitative by nature, and often focuses on one or two examples of controversy (c.f. \cite{szivos2005temporality,van2008controversy}), or else works towards a more qualitative analysis of the overall patterns across controversies \cite{Dascal1995}, with one notable exception \cite{Cramer2011}. In philosophy, Leibniz offered a simple definition of controversy: a controversy is a question over which contrary opinions are held \cite{leibniz1982vorausedition}, which Dascal notes as ``clearly insufficient'' \cite{Dascal1995}. Dascal offers a theory of controversies which distinguishes between types of polemic discourse \cite{Dascal1995}. Chen and Berger, while discussing whether controversy increases buzz and whether that is good for business, propose that ``controversial issues tend to involve opposing viewpoints that are strongly held'' \cite{Chen2013}. However, these definitions leave a gap when people disagree on opinions that are strongly held on frivolous topics such as the colors of a dress. 
We depart from past research by hypothesizing controversy as a multidimensional quantity, of which ``contention'' and ``importance'' are possible dimensions and which accounts for such differences.

Our model for contention draws on insights from existing computational, humanities and social sciences work, yet departs from it in important ways, and offers a re-conceptualization of controversy. 







\begin{table*}[t]
\caption{Data sets containing explicit stances
 }\label{tbl:datasetPolls}
\begin{center}
\resizebox{0.95\textwidth}{!}{
\begin{tabular}{ l l r l c r p{5.5cm} }
 Dataset & Type & \# Issues & Population(s) & Years & \# People & Source \\ \hline
 Gallup & Statistically Calibrated Phone Survey & 3 
     & US adults & 1939-2016 & varies (K) & \cite{deathGallup,gayGallup,drugsGallup} \\ 
 Pew Adults & Statistically Calibrated Phone Survey & 13 & US adults & 2014 & 2.0K & \cite{PewResearchCenter2015,PewResearchCenter2015a} \\  
 Pew AAAS & Statistically Calibrated Online Survey & 13 & US scientists & 2014 & 3.7K & \cite{PewResearchCenter2015,PewResearchCenter2015a} \\ \hline
 iSideWith & Informal Online Polling & 52 & US people & 2014 & varies (M) & By request \\
 Dress Buzzfeed & Informal Online Polling & 1 & Online readers & 2015-2016 & 3.5M & \cite{buzzfeedPoll} \\ \hline
 Brexit Votes & Public Voting Records & 1 & UK voters & 2016 & 46.5M & \cite{brexitData} \\
 U.S. Votes & Public Voting Records & 1 & U.S. voters & 2016 & 251.1M & \cite{McDonald2017,Wasserman2017,ElectionsWP2017}
\end{tabular}
}  
\end{center}
\end{table*}

\begin{table}[tb]
\caption{Twitter Data set with implicit stances
 }\label{tbl:datasetTweets}
\begin{center}
\resizebox{0.95\columnwidth}{!}{
\begin{tabular}{ l r r l }
Topic & \# Tweets & \# Users & Dates \\ \hline
\dress{} & 408.1K & 296.9K & Feb. 26-Mar. 9, 2015 \\
Brexit Referendum & 1.2M & 604.1K & May. 7-Aug. 24, 2016 \\ 
U.S. Elections & 87.4M & 10.1M & Sep. 20- Nov. 31, 2016 \\
Rio Olympics & 4.6M & 1.9M & Aug. 1-Aug.30, 2016 \\
Pokemon Go & 3.2M & 1.5M & Aug. 1-Aug.30,2016 \\
Nepal Earthquake & 49.8K & 36.3K & Apr.24-Apr.30,2015 \\ \hline
Total & 96.9M & 14.4M & \\

\end{tabular}
} 
\end{center}
\end{table}


\section{Modeling Contention}
\label{sec:modelC}

We now mathematically formulate a measure we call ``contention'', which quantifies the proportion of people in disagreement within a population. We begin with a general formulation of contention, and then describe a special case in which stances are assumed mutually exclusive.

Let $\Omega = \{p_1 .. p_n\}$ be a population of $n$ people. Let $T$ be a topic of interest to at least one person in $\Omega$. 

We define $c$ to denote the level of contention, which we define with respect to a topic and a group of people: $P(c| \Omega, T)$ represents the probability of contention of topic $T$ within $\Omega$. Let $P(nc| \Omega, T)$ similarly denote the probability of non-contention with respect to a topic and a group of people, such that: $P(c| \Omega, T) + P(nc | \Omega, T) = 1$.

Let $s$ denote a stance with regard to the topic $T$, and let the relationship $holds(p,s,T)$ denote that person $p$ holds stance $s$ with regard to topic $T$. 
Let $\hat S = \{s_1, s_2, .. s_k\}$  be the set of $k$ stances with regard to topic $T$ in the population $\Omega$. 
We allow people to hold no stance at all with regard to the topic (either because they are not aware of the topic, or they are aware of it but do not take a stance on it). We use $s_0$ to represent this lack of stance. In that case, let 
\begin{equation*}
holds(p,s_0,T) \iff \not\exists s_i \in \hat S \mbox{ s.t. } holds(p,s_i,T),
\end{equation*}

Let $S = \{s_0\} \cup \hat S$ be the set of $k + 1$ stances with regard to topic $T$ in the population $\Omega$. Therefore, $\forall p \in \Omega$, $\exists s \in S$ s.t. $holds(p,s,T)$. Now, let \textit{conflicts}: $S \times S \rightarrow \{0,1\}$ be a binary function which represents when two stances are in conflict. Note that a person can hold multiple stances simultaneously, though no stance can be jointly held with $s_0$. We set conflicts$(s_i,s_i)=0$. 

Let \textbf{stance groups} in the population be groups of people that hold the same stance: for $i \in \{0..k\}$, let $G_i = \{p \in \Omega | holds (p, s_i, T) \}$. 
By construction, $\Omega = \bigcup_i G_i$. Let \textbf{opposing groups} in the population be groups of people that hold a stance that conflicts with $s_i$. For $i \in \{0..k\}$, let $O_i = \{p \in \Omega | \exists j$ s.t. $holds (p, s_j, T) \land conflict(s_i, s_j) \}$. 

As a reminder, our goal is to quantify the proportion of people who disagree. Intuitively, we would like to have that quantity grow when the groups in disagreement are larger. In other words, if we randomly select two people, how likely are they to hold conflicting stances?

We model contention directly to reflect this question. Let $P(c| \Omega, T)$ be the probability that if we randomly select two people in $\Omega$, they will conflict on topic $T$. This is equal to:
\begin{multline*}
    P(c | \Omega, T) = P(p_1, p_2 \mbox{ selected randomly from } \Omega, \exists s_i, s_j \in S, \\ \mbox{ s.t. } holds(p_1, s_i, T) \land holds(p_2, s_j, T) \land conflicts(s_i,s_j))
\end{multline*}

Alternatively:
\begin{multline*}
    P(c | \Omega, T) = P(p_1, p_2 \mbox{ selected randomly from } \Omega, \exists s_i \in S, \\ \mbox{ s.t. } p_1 \in G_i \land p_2 \in O_i).
\end{multline*}

Finally, we extend this definition to any sub-population of $\Omega$. Let $\omega \subseteq \Omega, \omega \neq \emptyset$ be any non-empty sub-group of the population. Let $g_i = G_i \cap \omega$, and $o_i = O_i \cap \omega$. Thus, by construction, $g_i \subseteq G_i$ and $\omega = \bigcup_i g_i$. The same model applies respectively to the sub-population. In other words, for any $\omega \subseteq \Omega$,
\begin{align*}
P(c | \omega, T) = & P(p_1, p_2 \mbox{ selected randomly from } \omega 
                \\ & \land \exists i \mbox{ s.t. } p_1 \in g_i \land p_2 \in o_i).
\end{align*}

\subsection{Mutually exclusive stances}

Note that we are selecting with replacement, and it is possible for $p_1=p_2$. Strictly speaking, this model allows a person to hold two conflicting stances at once and thus be in both $G_i$ and $O_i$, as in the case of intrapersonal conflict.
This definition, while exhaustive to all possible combinations of stances, is very hard to estimate. We now consider a special case of this model with two additional constraints. 
Let every person have only one stance on a topic: 
\begin{equation}
\begin{split}
\not\exists p \in \Omega, s_i, s_j \in S & \mbox{ s.t. } i \neq j \land 
                 \\ & holds(p,s_i,T) \land holds(p,s_j,T).
\end{split}
\end{equation}
And, let every explicit stance conflict with every other explicit stance:
\begin{equation}
\mbox{\textit{conflicts}}(s_i,s_j) \iff (i \neq j \land i \neq 0 \land j \neq 0)
\end{equation}
This implies that $G_i \cap G_j = \emptyset$. Crucially, we set a lack of stance not to be in conflict with any explicit stance. Thus, $O_i = \Omega \setminus G_i \setminus G_0$.

For simplicity, we estimate the probability of selecting $p_1$ and $p_2$ as selection with replacement\footnote{The calculation is very similar for selection without replacement, except for extremely small population sizes.}. 
Note that $|\Omega| = \Sigma_{i \in \{0..k\}} |G_i|$ and the probability of choosing any particular pair is $\frac{1}{|\Omega|^2}$. The denominator, $|\Omega|^2$, expands into the following expression:
\begin{multline*}
|\Omega|^2 = (\Sigma_i |G_i|)^2 
= \Sigma_{i \in \{0..k\}} |G_i|^2 + \Sigma_{i \in \{1..k\}} (2 |G_0| |G_i|) \\ + \Sigma_{i \in \{2..k\}} \Sigma_{j \in \{1..i-1\}} (2 |G_i| |G_j|)
\end{multline*}
Depending on whether the pair of people selected hold conflicting stances or not, they contribute to the numerator in $P(c|\Omega,T)$ or $P(nc|\Omega,T)$, respectively. Therefore,
\begin{equation*} P(c|\Omega,T) = \frac{\Sigma_{i \in \{2..k\}} \Sigma_{j \in \{1..i-1\}} (2 |G_i| |G_j|)}{|\Omega|^2} \end{equation*}
and
\begin{multline*}
P(nc|\Omega,T) = 1 - P(c|\Omega,T) = \\ \frac{\Sigma_{i \in \{0..k\}} |G_i|^2 + \Sigma_{i \in \{1..k\}} (2 |G_0| |G_i|)}{|\Omega|^2}
\end{multline*}

As before, we can trivially extend this definition to any non-empty sub-population $\omega \subseteq \Omega$ using $g_i = G_i \cap \omega$. By construction, there is no contention within any single-stance group, $g_i$, with respect to topic $T$. In other words, $P(c | g_i, T) = 0$. Additionally, by construction, there is no contention within $g_i \cup g_0$, i.e. $P(c | g_i \cup g_0, T) = 0$.

By extension, if there is only one explicit stance $s_1$ with regard to topic $T$ in the population $\Omega$, there will be no contention in the population with respect to the topic. In other words, $|\hat S| \leq 1 \implies P(c | \Omega, T) = 0$.

Trivially, $P(C|\omega,T)$ is maximal when when $|g_0| = 0$ and $|g_1| = ... = |g_k| = \frac{|\omega|}{k}$, and its value is $\frac{k-1}{k}$. This is subtly different from entropy due to the existence of $s_0$, as entropy would be maximal when $|g_0| = |g_1| = ... = |g_k| = \frac{|\omega|}{k-1}$.

Since the values of contention are $[0,\frac{k-1}{k}]$ rather than $[0,1]$, we normalize by the maximal contention (divide the contention score by $\frac{k-1}{k}$) and take the non-contention score as 1 minus the new score. This normalization brings both contention and non-contention to a full range of $[0,1]$ each, with a contention score of 1 signifying the highest possible contention, regardless of the total number of stances.

\begin{figure}[t]
	\includegraphics[width=0.9\columnwidth]{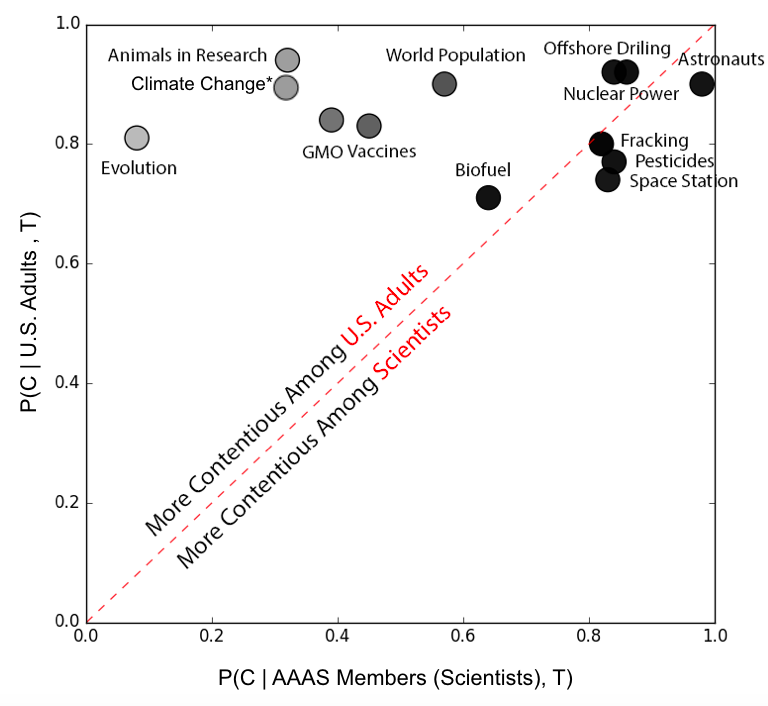}
	\caption{Contention in the scientific community vs.~general population for several controversial topics. \small{The x=y line represents equal contention among both populations, with dots shaded according to their distance from the line. Note that the Climate Change question had 3 explicit stances, all other questions had 2. 
	}}
    \label{fig:AAAS}
\end{figure}

\begin{table}[]
\centering
\caption{Examples of questions for the topics in Figure \ref{fig:AAAS}\cite{PewResearchCenter2015,PewResearchCenter2015a} \small{(bold keywords match point labels).}}
\label{fig:AAAS_topics}
\resizebox{0.95\columnwidth}{!}{
\begin{tabular}{@{}l@{}}
\multicolumn{1}{c}{\textbf{Issues}}                      \\
\toprule
\textbf{Q:} \small{Opinion on the increased use of \textbf{fracking}:} \\
\textbf{A:} \small{\{Favor, Oppose\}} \\ \hline
\textbf{Q:} \small{The \textbf{space station} has been ... for the country:} \\  
\textbf{A:} \small{\{Good investment, Not a good investment\}} \\ \hline
\textbf{Q:} \small{Thinking about childhood diseases, such as measles, mumps,} \\ 
\small{rubella and polio, do you think... (\textbf{label: ``vaccines''})} \\
\textbf{A:} \small{\{All children should be required to be vaccinated, Parents } \\ 
\small{should be able to decide NOT to vaccinate their children\}} \\ \hline
\textbf{Q:} \small{Do you think it is generally ... to eat foods grown with} \\
\small{\textbf{pesticides}}. \textbf{A:} \small{\{Safe, Unsafe\}} \\ \hline
\end{tabular}
} 
\end{table}

\begin{figure}[t]
\begin{center}
	\includegraphics[width=0.95\columnwidth]{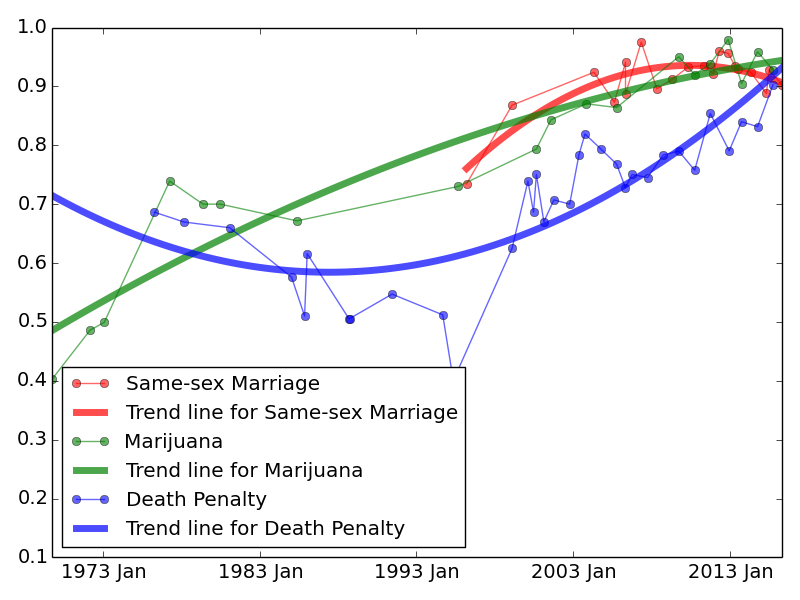}
	\caption{Contention over time for three controversial topics. \small{Trendlines are 2nd degree polynomials. Results for ``Death Penalty'' prior to 1969 are omitted.}}
    \label{fig:gallupTime}
\end{center}
\end{figure}

\section{Data Collection and Preparation}


In order to ground our model in empirical data, we collected several data sets. First, we collected data sets that represent explicit stance information, from informal online polls, through phone surveys, to actual voting records (on Brexit and the 2016 U.S. Elections). The complete set of explicit-stance data sets appears in Table \ref{tbl:datasetPolls}. Between these data sets, we cover a wide variety 
of public opinion issues and a span of over 50 years. 
Second, we collected a set of tweets on several topics, one focusing on Brexit, and the other on ``The Dress'' phenomenon (see Table \ref{tbl:datasetTweets});  in both, the stances taken by people are implicit and must be estimated.

\subsection{Polling data sets}

In the Pew and Gallup data sets, we used the topline survey results as reported by the respective organizations. For a given poll topic $T$, $\omega$ is the set of respondents, $s_i$ are the set of response possibilities, and ``no answer'' represents $s_0$. This determines $g_i$ and thus allows us to calculate $P(c|\omega,T)$ as above. In the case of statistically representative polls, conclusions can be generalized for the wider population from which the poll sample was drawn (within the margin of error of the polls).

Using one data set acquired from Pew Research Center, a non-partisan fact tank in the U.S., we are able to examine attitudes towards a number of issues among two populations: U.S. adults and U.S. scientists (Pew Adults and Pew AAAS in Table \ref{tbl:datasetPolls}). The opinions for U.S. adults was gathered among a representative sample of 2,002 adults nationwide, while the opinions for scientists were gathered among a representative sample among the U.S. membership of the American Association for the Advancement of Science (AAAS) \cite{PewResearchCenter2015a}.

We also obtained a data set from the iSideWith.com website, a nonpartisan Voting Advice Application~\cite{Cedroni2010} which offers users the chance to report their opinions on a wide variety of controversial topics, and outputs the information of which political candidate they most closely align with. We received the 2014 iSideWith data set by request from the website owners, which included nation-wide and per-state opinions over 52 topics. Each topic was posed as a question with two main options for answers, usually simply ``yes'' and ``no''. Additionally, the data set included the average importance of the issue (both nation-wide and per-state) rated by the users, which we use in our hypothesized controversy model (but not for contention).

\subsection{Twitter data set}

We collect a set of tweets on six events or topics from Twitter, which is available on our website\footnote{
\url{http://ciir.cs.umass.edu/irdemo/contention/dataset/}}.
We selected three contentious topics: \dress{}, the Brexit referendum, and the 2016 U.S. elections. 

From the collected tweets, we identify two sub-groups of tweets by their stance revealed through their hashtags in order to measure their contention. In addition to the Twitter data, we also collected actual voting records for Brexit and the U.S. elections (see below for further description), as well as the Buzzfeed poll results for \dress{} (see Table \ref{tbl:datasetPolls}). For this purpose, we use the Twitter Garden Hose API, which allows us to collect 10\% random sample of actual tweets if it is included in the sample.


\textbf{\dress{}} refers to a photo that went viral over social media starting Feb.~26, 2015, after people couldn’t agree on its colors. The photo was posted to tumblr and made popular by a Buzzfeed article asking “What color is this dress?” as a poll with two options, black and blue or gold and white; over 37 million people viewed the article to date \cite{buzzfeedPoll}. Over the course of the next 24 hours, \dress{} made headline news in mainstream media outlets. The actual dress was discovered to be black and blue, but the surprising photo continues to be a source of exploration for scientists of vision perception \cite{dressSpecialIssue}. For this topics, we collected tweets that contain relevant hashtags from the Garden Hose API. We used four popular hashtags as seeds, \textsc{\#dressgate, \#thedress, \#whiteandgold,} and \textsc{\#blackandblue}, then extracted the frequent hashtags from the collected tweets and manually verified those relevant to \dress{} (Table \ref{tbl:hashtags}). We then generated two groups of hashtags, each of which represents one of two stances: seeing the dress as white and gold, and seeing the dress as black and blue. Among the hashtags in the collected tweets, we extracted one set of hashtags that contain both ``black'' and ``blue'', and the other set that contain ``white'' and ``gold''. We also extracted comparable hashtags in multiple languages, using a list of those color names translated into 80 different languages. We retrieved hashtags that contained the translated words for both ``black'' and ``blue'' or both ``white'' and ``gold'' in the same language, such as \textsc{\#negroyazul}\footnote{`Negro' means `black' in Spanish and `azul' means `blue'.}. 

\textbf{The Brexit referendum,} officially known as the United Kingdom European Union membership referendum, was a referendum that took place on June 23, 2016 in which 51.9\% of UK voters voted to leave the EU. While not legally binding, the referendum had immediate political and financial consequences, including the worst one-day drop in the worldwide stock market in history to that date, and the resignation of then-Prime Minister David Cameron.

For Brexit, we downloaded a set of tweet ids\footnote{\url{http://www.eecs.qmul.ac.uk/~dm303/brexit}} collected and released by Milajevs using the tool Poultry \cite{Milajevs2013RealTD}. The data set contained tweet ids related to Brexit from March 7 to August 24th. For each tweet id, we retrieve the corresponding tweet via Twitter Garden Hose API, which allows us to collect 10\% random sample of actual tweets if it is included in the sample. Through this process, we were able to obtain 1,222,313 tweets, which is 5.2\% of the released Tweet ids for Brexit. Then we used manually curated hashtags to find two stance groups of the tweets, if any stance is revealed in the tweet (Table \ref{tbl:hashtags}).

\textbf{The 2016 U.S. Presidential Elections} were widely considered one of the most rancorous elections in recent U.S. history, and attracted not only U.S. but also worldwide attention. The two major conflicting stances were with regards to the two main presidential candidates, Donald Trump and Hillary Clinton. To observe the contention trend before, during, and after the voting day, we collected tweets that contain election-related hashtags from Sep 20, 2016 to Nov 30, 2016. We start from the straightforward topic hashtags such as \{\#election2016, \#presidentialelection, \#hillaryclinton, \#donaldtrump\} and a few keywords such as \{president, election, hillary clinton, donald trump\} as seeds. Tweets are collected if they contain any of the predefined topic hashtags or keywords. From the collected tweets, we look at the top 50 frequent hashtags and extend the seed hashtag set by adding other relevant hashtags. 

To detect the stances, we extracted the top 50 frequent hashtags from the collection. Three expert annotators annotated whether a given hashtag explicitly indicates a stance on which presidential candidate the tweet supports, and we selected only hashtags that all annotators agreed on. Some hashtags contain stances to some extent, but the stances can be either way depending on the context such as \textsc{\#hillarybecause} and \textsc{\#draintheswamp}. To take a high-precision, rather than low-recall approach, we extract the set of stance hashtags that three annotators agreed on (Table 4).

In all three cases, we use a high-precision, low-recall approach to detect stances by only assigning a stance to tweets that use an explicit stance hashtag, such as \textsc{\#blackandblue} or \textsc{\#leaveeu}. We release a complete list of hashtags used on our website, along with the tweet ids for the collection. While we are certain to miss a large portion of stance-taking tweets that do not use these hashtags, this allows us to be reasonably confident that the stances detected are accurate, which is most useful for the purposes of model validation. We leave analysis of the remaining tweets and other hashtags for future work in stance extraction.

Using the stance hashtags we created, we compute the size of the two stance groups per topic by counting the number of tweets that contain any hashtag from each stance. As an estimation of $G_0$ (the group with no stance) on each topic, we used all other tweets collected via the Twitter Garden Hose API that day. Specifically, $|G_0|$ = count of all tweets collected $- |G_1| - |G_2|$. 

\textbf{Non-controversial topics.} In order to validate our model on a range of topics, we also collected Twitter data for three prominent and essentially non-controversial topics: The mobile game Pokemon Go, The 2016 Rio Olympics, and the 2015 Nepal Earthquake. For each of these topics, we examined the top 30 frequent hashtags to check if there exists any conflicting stance. We did not expect to find any conflicting stances in these hashtags, and a close examination of the top 30 hashtags confirmed this. We therefore omit further analysis of these topics for this paper.

\subsection{Voting data for Brexit and U.S. Elections}


We collected actual voting data for Brexit and the 2016 U.S. Elections. The Brexit voting data, including turnout figures, was released by the UK Electoral Commission \cite{brexitData}, and was split by Unitary Districts. The EU referendum only had two options, ``Remain'' or ``Leave'', which represent two conflicting stances. We considered any non-voters or rejected ballots as having no stance.

For the U.S. Elections, the Federal Election Committee has not released its official results by the time of writing. Nonetheless, we were able to collect the election results from two sources. We used the Popular Vote Tracker \cite{Wasserman2017} for certified state results on the 2 major candidates, Donald Trump and Hillary Clinton. Additionally, we used results tabulated on Wikipedia \cite{ElectionsWP2017}, which at the time of writing were official in all but 2 states; these figures included a break-down of results for the three main third party candidates (Johnson, Stein and McMullin) and ``Other''. Estimated turnout figures were collected from the Elections Project \cite{McDonald2017}; we used the reported VEP Highest Office turnout metric, which is available for all U.S. states, to estimate the amount of people holding no stance.  

\section{Model Validation}  

In order to ground our model and ensure that it aligns with actual controversies, we use our model to measure contention on the data sets described above. 

\begin{figure}[t]
\begin{center}
	\includegraphics[width=0.9\columnwidth]{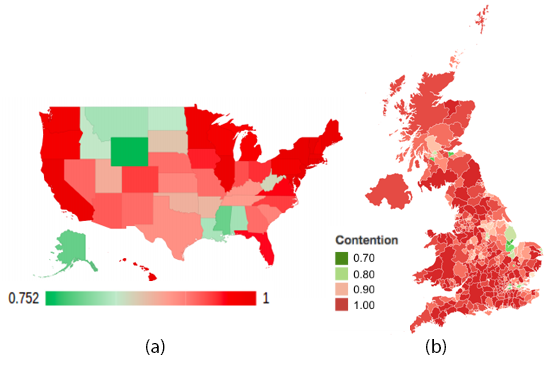}
	\caption{(a) Per-state contention for ``Do you support increased gun control?". (b) Contention by voting district in the UK (The Electoral Comission 2016) \small{Interactive maps for all iSideWith issues are available at 
	http://ciir.cs.umass.edu/irdemo/contention/isidewith/
	.}}
    \label{fig:iSideWith}
\end{center}
\end{figure}

\subsection{Contention in Polling}

We first validate our model using the polls in Table 
Table \ref{tbl:datasetPolls}).  We describe a few patterns that emerge.

\textbf{U.S. Scientists vs.~General Population.} 
Using the Pew Research data sets (Pew Adults and Pew AAAS in Table \ref{tbl:datasetPolls}), we are able to examine attitudes towards a number of scientific issues among two populations: U.S. adults and U.S. scientists.

 As seen in Figure \ref{fig:AAAS}, for some topics such as offshore drilling, hydraulic fracturing (fracking), and biofuel, contention was similar between U.S. adults and scientists. On other topics, such as evolution, climate change, and the use of animals in research, contention varied widely depending on the population: the scientific community had low contention for these topics, whereas they were highly contentious among U.S. adults. This result precisely matches prior work's intuitive notion of politically, but not scientifically, controversial topics \cite{Wilson2015}. The graph clearly demonstrates the notion that ``there is no controversy'' (among scientists) alongside the controversy in general population, with evolution as the most extreme case presented in this data set (98\% of AAAS members surveyed said that ``humans and other living things have evolved over time'', whereas 31\% of the U.S. adults said that they have ``existed in their
 present form since beginning of time''). 


\begin{table*}[]
\centering
\caption{\small{Example hashtags used to identify two stance groups on \dress{}, Brexit and the U.S. Elections. Full list at \demo{}.}}
\label{tbl:hashtags}
\resizebox{0.95\textwidth}{!}{
\begin{tabular}{@{}llll@{}}
\toprule
\multicolumn{1}{c}{\textbf{Topic}} & \multicolumn{1}{c}{\textbf{Stances}} & \multicolumn{1}{c}{\textbf{Example Hashtags}} & \multicolumn{1}{c}{\textbf{\# of hashtags}} \\ \midrule
\multirow{2}{*}{The Dress} & Blue and Black &
\#blackandblue, \#notwhiteandgold,  \#blackandbluedress,\#\begin{CJK}{UTF8}{min}青と黒\end{CJK},\#negroyazul ... & 49  \\
& White and Gold      &    \#whiteandgold, \#whiteandgoldteam, \#thedressiswhiteandgold,\#blancodorado ...   & 37  \\
\multirow{2}{*}{Brexit} & Leave EU                             & \#voteleave, \#leave, \#leaveeu, \#betteroffout    & 4                                            \\
    & Remain EU  & \#remain, \#strongerin, \#voteremain,  \#regrexit, \#remainineu     & 5    \\
\multirow{2}{*}{U.S. Election}     & Hillary Clinton     &    \#imwithher, \#strongertogether, \#dumptrump, \#notmypresident ...          & 10                                            \\
     & Donald Trump     & \#maga, \#trumppence, \#trumptrain ...   &  26                                           \\ \hline
\end{tabular}
}
\end{table*}


\textbf{Contention over time for ``hot button'' topics.} The Gallup data set gives us access to changing contention over time for several controversial topics in the U.S. We selected three topics: the death penalty for murder, legalization of marijuana, and legalization of same-sex marriage. As seen in Figure \ref{fig:gallupTime}, clear trends emerge when contention is mapped over time. For example, marijuana legalization had consistently low contention in the early '70s (when less than 20\% of the population thought it should be legalized); support for the death penalty was high (and contention low) during the '90s. Interestingly, contention for both marijuana legalization and same-sex marriage peaked recently, and is now going down as the support for each of these has crossed the threshold of 50\% around 2012. For the death penalty, contention between sub-populations in the U.S. varied widely; for example, contention was higher among the black and Hispanic populations, and higher for democrats (full results omitted for space considerations).


\textbf{Per-state distribution of Contention in the United States.} 
Using the iSideWith data set, we measured contention nation-wide and per-state on each of the 52 topics available. The two least contentious questions nation-wide were ``Should National Parks continue to be preserved and protected by the federal government?'' ($P(c|US,t)$ = 0.26), and ``Should every person purchasing a gun be required to pass a criminal and public safety background check?'' ($P(c|US,t)$ = 0.39). Several topics had over 0.99 contention nation-wide, such as ``Should the U.S. formally declare war on ISIS?'' and ``Would you support increasing taxes on the rich in order to reduce interest rates for student loans?", among others. We present the per-state contention for one such topic in Figure \ref{fig:iSideWith}, which shows how contention varies geographically. An interactive demo with per-state contention on all 52 topics is available at \url{http://ciir.cs.umass.edu/irdemo/contention/isidewith/}. 


\begin{figure*}[t]
\begin{center}
	\includegraphics[width=\textwidth]{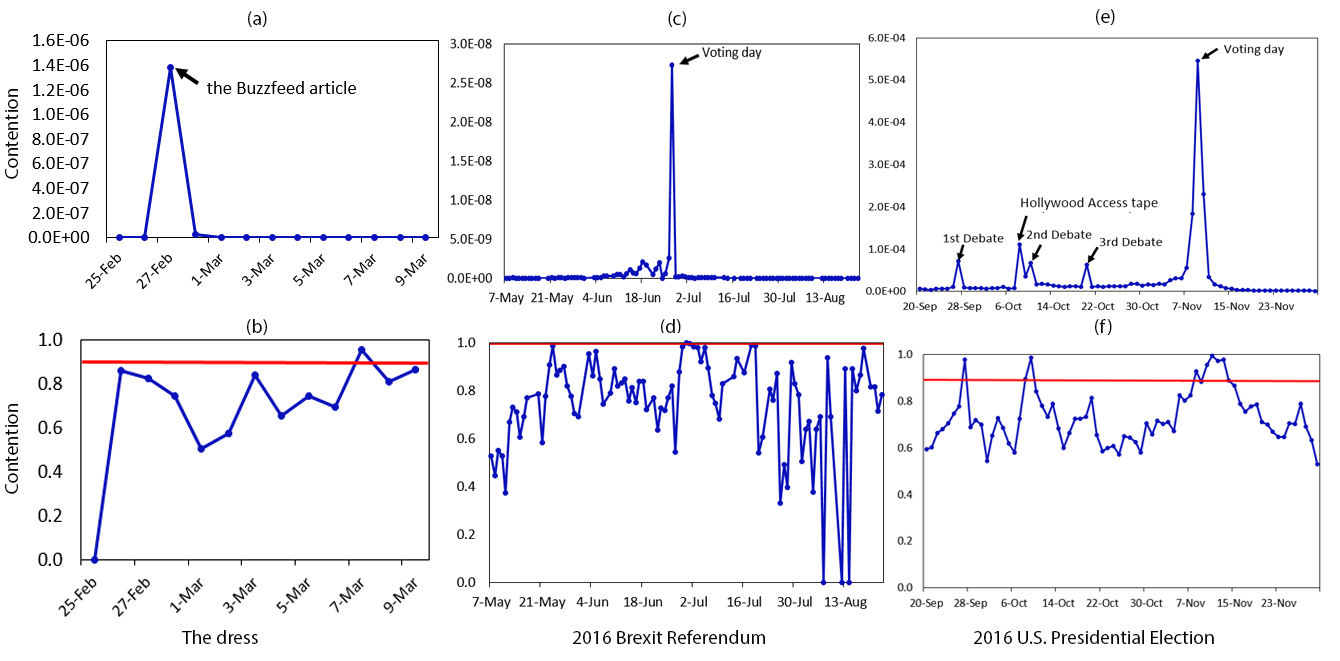}
	
	\caption{Contention among all daily tweets by date for \dress{} (left), Brexit (center) and 2016 U.S. Elections (right), reported among all Gardenhose tweets that day (top) or only among those with an explicit stance (bottom). Notable peaks are annotated with associated events around that time. All dates are in UTC. The horizontal lines in (b), (d), (f) show the contention from alternate sources (\dress{}, 0.88; Brexit, 1.00; U.S. Elections, 0.89).}
	\label{fig:contentionTwitter}
\end{center}
\end{figure*}

\subsection{Contention on Twitter} 
\label{Sec:TheDressBrexit}

From the Twitter data collected above, we report contention for our three controversial topics: \dress{}, Brexit and the U.S. elections. For each topic, we calculate two types of daily contention trends: one, only among the tweets exhibiting a stance on the topic on that day, and the other among all of the Twitter posts on that day, i.e., including $G_0$. A visible pattern emerges, where contention only among the population that exhibits a stance is consistently high throughout, whereas including $G_0$ shows marked peaks of contention around notable event times. For example, in the U.S. Elections case, small peaks appear on the days of the presidential debates, and upon release of the extremely controversial Hollywood Access tape, with a much larger peak on election day. This showcases the strength of our model and its ability to track the difference between contention among the group for which the topic is salient ($G_1 \cup G_2$), as opposed to the entire population. 

\textbf{Comparison to external sources.} We compare $P(c|G_1 \cup G_2,T)$ from Twitter across a series of dates, with that calculated from external sources: the Buzzfeed poll on \dress{} ($P(c|G_1 \cup G_2,T)$ = 0.88) \cite{buzzfeedPoll}, voting results on Brexit ($P(c|G_1 \cup G_2,T)$ = 1.00) \cite{brexitData}, and the popular vote in the U.S. Elections measured for the two main candidates ($P(c|G_1 \cup G_2,T)$ = 0.89). Additionally, Figure \ref{fig:iSideWith}(b) shows the voting contention for each Unitary District of the UK (local Ireland results were not available), demonstrating the geographical variance of contention. Gibraltar, an extreme outlier both geographically and contention-wise, is omitted from the map ($P(c|Gibraltar,Brexit) = 0.16$). The extremely low contention makes sense: Gibraltar is geographically located inside Europe, and 95.9\% of its voters voted ``Remain''.

\textbf{Turnout in voting}. For the 2016 United States elections and Brexit, we measured contention with or without estimated turnout figures. In both cases, $G_0$ was set as the number of eligible voters (official in the UK, estimated in the U.S.) who did not vote. Contention decreases markedly when voter turnout is factored into the model. For the extremely divisive U.S. elections, contention dropped from 0.89 to 0.31 when factoring in the estimated 41.1\% of eligible voters that did not go to the ballots on election day. A similar pattern is observed for Brexit.


\textbf{Contention and Third-Party Votes.} We briefly analyzed the results of contention in the U.S. Elections as measured on the two main candidates as well as the three main third-party candidates, Johnson, Stein and McMullin and a sixth category reported as ``Other'' \cite{ElectionsWP2017}. A few interesting patterns are revealed when examining this six-way contention. For example, measured only on Trump and Clinton, contention is nearly the lowest in Utah, but is highest of all states when considering the third party candidates. This makes sense when considering that Evan McMullin received 21.3\% of the vote in that state. 

\section{Reconceptualizing Controversy}

In light of our new measure of \textbf{contention}, we now reexamine the idea of \textbf{controversy} and hypothesize a model for it, which should likewise be based on a notion of the population observed. We suggest an approach which rather than modeling controversy directly, contains multiple dimensions contributing to controversy, with contention being one of them. A certain level of contention may or may not meet criteria for controversy, depending on other dimensions of the controversy model. 

As before, $\Omega = \{p_1 .. p_n\}$ is a population of $n$ people, and $T$ is a topic of interest. We thus define the level of controversy with respect to a topic and a group of people: Let $controversy(\Omega, T)$ represents the level of controversy of topic $T$ within $\Omega$. 

\begin{figure}[t]
	\includegraphics[width=\columnwidth]{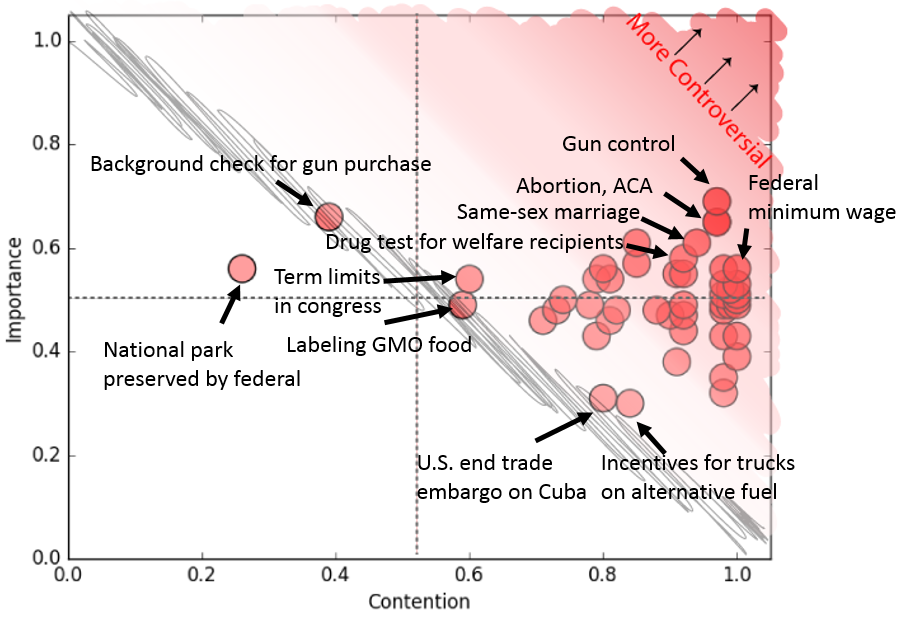}
	\caption{iSideWith topics plotted reconceptualizing controversy as composed of at least two dimensions, contention and importance. Sample topics are given in each quadrant of \{low,high\} importance and contention.}
    \label{fig:multidim}
\end{figure}

\subsection{Controversy is Multidimensional}
\label{sec:multidim}

Consider the cases of the Brexit referendum and \dress{}, two controversies which we explored in detail above.
When observed among the population which considered them as salient, both were extremely contentious in the sense that nearly any group of people sampled from these populations was strongly divided in their opinion. However, it is immediately obvious that placing Brexit and \dress{} in the same bucket is somewhat problematic.
One, a political referendum on Britain's decision whether to exit the European Union, affects the fate of entire nations, with far-reaching and difficult to predict effects on diplomatic relationships and the world economy for years to come. 
The other, an under-developed photo of a mother-of-the-bride's dress, caused a surprising divided reaction in color perception, went viral around the world, and was subsequently forgotten by nearly everyone. Its impact on the world was likely negligible, with the exception of a burst of scientific papers in visual perception studying this unexpected effect~\cite{Schlaffke2015,dressSpecialIssue}\footnote{And now, this paper.}.

Therefore, we propose a new model in which controversy is composed of at least two orthogonal dimensions, which together play a role in determining how controversial a topic is for a given population, one of which is ``contention''. We can hypothesize other possible dimensions. For example, a possible second dimension is ``conviction'', i.e. encoding how strongly people hold their opinions \cite{Chen2013}. However, this dimension is insufficient to explain such arguably frivolous controversies as \dress. An additional orthogonal metric is needed in order to distinguish between contention and controversy. Therefore, we hypothesize the existence of a notion of ``importance" or ``impact'' as a novel dimension of controversy. Using the same notation as above, we hypothesize that these are minimal dimensions of controversy, though there may be others:
\begin{align*}
controversy(\Omega, T) = f( & contention(\Omega, T), \\ & conviction(\Omega, T), \\ & importance(\Omega, T)...)
\end{align*}

This framework is demonstrated schematically in Figure \ref{fig:multidim}, overlaying actual results including importance reported in the iSideWith data set (see Table \ref{tbl:datasetPolls}). The first dimension is ``contention'' which we defined as the proportion of people who are in disagreement. The other dimension is ``importance'', which we loosely define as the level of impact of that issue to the world, and which was reported by users of iSideWith. In Figure 5, we hypothesize controversy to be a two-dimensional concept. An issue is more controversial when it has high contention and high importance (i.e., towards right upper corner of Figure 5). Figure 5 shows a quadrant where an issue can have a \{high, low\} contention with a \{high, low\} importance. Issues such as gun control, abortion, and affordable care act have high contention and high importance, hence more controversial. Issues such as whether the government should provide incentives for trucks to run on alternative fuels is highly contentious but is rated by users as low importance. Likewise, whether National parks should be preserved by the federal government is considered rated as somewhat important, but not contentious. 
Using this framework, we can understand the disparity between \dress{} and Brexit: the former is contentious with low importance, and thus not as controversial as Brexit with its high contention and high importance.


While computationally exploring the additional hypothesized dimensions, ``conviction'' and ``importance'' is beyond the scope of this paper, we have demonstrated that contention clearly is one such dimension, and that at least one additional dimension is required in order to fully understand controversies; contention does not fully capture the nuances of what we intuitively understand to be controversial. 


\section{Discussion}

Our population-based contention model offers a new way of quantifying controversies, and a new way to understand multiple observed phenomena, only some of which we covered in this paper. For example, an conflict between two populations will often have low contention internally since each is fairly consistent with a specific stance, but when the two populations are observed together, the combination is highly contentious. Small, community-specific controversies can now be quantified as well; a certain topic might be extremely controversial in a tight-knit population, 
while the rest of the world is starkly in $G_0$, either oblivious or apathetic to the controversy. Other population-dependent contention levels can be observed elsewhere, for example in the case of racial tensions around police brutality in the U.S. 
As demonstrated in Figure \ref{fig:AAAS}, we can use this model to quantify the aforementioned high-stakes public opinion controversies over scientifically well-understood phenomena. To refer back to the question in our paper title: ``Is Climate Change controversial?'', the answer is: it depends on the population being observed. In the scientific community, this and certain other topics such as evolution and vaccines are in consensus, while in the general U.S. population, their contention remains high.

For the purpose of model validation, we intentionally chose to use a high-precision, low-recall manual curation process to classify stances. However, we note that this high-quality curation is not central to the contention model: implicit or inferred stances can be used in the same manner. In fact, this stance detection process can be automated, as demonstrated by recent work \cite{coletto2016polarized,Garimella2016}, and such advances are synergistic with our contention metric. 

\subsection{Model Limitations}

As noted in Section \ref{sec:modelC}, our model allows for overlapping stances which are in practice very challenging to estimate. The added constraints of mutually exclusive stances which all conflict equally make the model extremely practical and easy to estimate; however, one must take care to ensure that the stances fed to the model are indeed mutually exclusive, otherwise the conclusions may not hold. The constraints are certainly true for many controversial topics, but not all of them. For example, for \dress{} we know there was a subset of people who in practice saw both color combinations, which we did not take into account.
Even for mutually-exclusive stances, comparison between issues with a varying number of stances may be complicated by the normalization factor, and further exploration is needed to understand this effect better. Additionally, if multiple stances lie on a spectrum between two extremes, it does not make sense to consider them all equally conflicting. In such a case, recasting the \emph{holds} and \emph{conflicts} functions to return a real value in the [0,1] range instead of a binary value may be a better fit; such a ``variable edit distance'' function is well known in the bioinformatics space, and existing work in that space could be leveraged for contention. Such a recasting might result in a more nuanced characterization of multi-stance controversies and allow a better comparison between them and two-sided controversies. We leave these analyses for future work.

\subsection{Future work} 

Our theoretical model of contention points the way to several possible avenues of future research. 
As mentioned above, stance extraction is a growing research topic \cite{coletto2016polarized,Garimella2016}, and automated stance extraction can certainly be applied to improve the detection and measurement of contention in the near future. 
An alternative conception of contention could conceivably start from groups rather than individuals, in a model which would explain stance as a conclusion of group membership \cite{kahan2015climate}. The differentiation between overlapping and mutually exclusive stances might be useful for classification of controversiality, reminiscent of a recent partitioning approach to measuring controversy \cite{Garimella2016}. 


Our work also clearly calls out the need for more research into the additional dimensions of controversy, beyond contention. For example, ``importance'' as a dimension of controversy allows for further examination. Alternative dimensions that might contribute to our hypothesized controversy model, which we have yet to explore, include notions of ``conviction'' (how likely is a person to change their stance?), ``identity-centrality'' (how central is this controversy to the individual's identity?), as well as ``loudness'' or ``influence'': all people are considered equally when evaluating contention, when in fact the stances of certain ``thought leaders'' may have a disproportionate impact by increasing the diffusion of their stances.


\section{Conclusions}

Drawing on work from a variety of disciplines, we propose a new measure, contention, which mathematically quantifies the notion of ``the proportion of people disagreeing on this topic'' in a population-dependent fashion. By allowing contention to be evaluated in various sub-popu\-la\-tions, on topics with multiple stances, and on people holding no stance, we can quantify a wide variety of phenomena, such as the difference between scientific controversies and political ones, the change in contention over time, and local or cultural patterns in contention. This allows us, for example, to formally answer the question in the title of our paper, ``Is Climate Change Controversial?'', differently depending on the population being observed: climate change is not contentious in the scientific community, yet is in the general U.S. public. We validate our theoretical model on a wide variety of data sets from both off- and online sources, ranging from large informal online polls and Twitter data, through statistically calibrated phone surveys, and actual voting records. 

Finally, we hypothesize a new theoretical model for reconceptualizing controversy. We redefine controversy, like contention, as rooted in populations, and as multi-dimensional rather than a single quantity. We posit that contention is one such dimension, and present preliminary evidence that importance is another possible dimension. Our contention measure and the hyopthesized controversy model hold significant promise in offering a deeper understanding of the nature of controversies, increasing the likelihood of reproducibility of future work, and holding implications for social science, humanities and computer science research on controversies, with civic, social and science-communication implications.

\textbf{Acknowledgements}.
\small{The authors thank Marc-Allen Cartright, Jeff Dalton, Shay Hummel, Kiran Garimella, Seth Goldman, Justin Gross, Daniel Mishori, Brendan O'Connor, and Alena Vasilyeva for fruitful conversations and for pointing us to valuable resources. Special thanks to Taylor Peck and Nick Boutelier for providing us the iSideWith data set. This work was supported in part by the Center for Intelligent Information Retrieval and in part by NSF grant \#IIS-1217281. Any opinions, findings and conclusions or recommendations expressed in this material are those of the authors and do not necessarily reflect those of the sponsor.}

\bibliographystyle{aaai}
\small\bibliography{referencesICWSM}

\end{document}